\documentclass[tighten,twocolumn]{aastex63}

\usepackage{graphicx}
\usepackage{ulem}
\usepackage{amsmath}
\usepackage{wrapfig}
\usepackage[thinlines]{easytable}
\usepackage{tikz}

\newcolumntype{P}[1]{>{\centering\arraybackslash}p{#1}}

\begin{document}
\shorttitle{FRBs in Globular Clusters}
\shortauthors{}

\title{Dynamical Formation Channels for Fast Radio Bursts in Globular Clusters}

\correspondingauthor{Kyle Kremer}
\email{kkremer@caltech.edu}

\author[0000-0002-4086-3180]{Kyle Kremer}
\altaffiliation{NSF Astronomy \& Astrophysics Postdoctoral Fellow}
\affiliation{TAPIR, California Institute of Technology, Pasadena, CA 91125, USA}
\affiliation{The Observatories of the Carnegie Institution for Science, Pasadena, CA 91101, USA}

\author[0000-0001-6806-0673]{Anthony L. Piro}
\affiliation{The Observatories of the Carnegie Institution for Science, Pasadena, CA 91101, USA}

\author[0000-0001-7931-0607]{Dongzi Li}
\affiliation{TAPIR, California Institute of Technology, Pasadena, CA 91125, USA}

\begin{abstract}
    The repeating fast radio burst (FRB) localized to a globular cluster in M81 challenges our understanding of FRB models. In this \textit{Letter}, we explore dynamical formation scenarios for objects in old globular clusters that may plausibly power FRBs.
    Using N-body simulations, we demonstrate that young neutron stars may form in globular clusters at a rate of up to $\sim50\,\rm{Gpc}^{-3}\,\rm{yr}^{-1}$ through a combination of binary white dwarf mergers, white dwarf--neutron star mergers, binary neutron star mergers, and accretion induced collapse of massive white dwarfs in binary systems. 
    We consider two FRB emission mechanisms: First, we show that a magnetically-powered source (e.g., a magnetar with field strength $\gtrsim10^{14}\,$G) is viable for radio emission efficiencies $\gtrsim10^{-4}$. This would require magnetic activity lifetimes longer than the associated spin-down timescales and longer than empirically-constrained lifetimes of Galactic magnetars. Alternatively, if these dynamical formation channels produce young rotation-powered neutron stars with spin periods of $\sim10\,$ms and magnetic fields of $\sim10^{11}\,$G (corresponding to spin-down lifetimes of $\gtrsim10^5\,$yr), the inferred event rate and energetics can be reasonably reproduced for order unity duty cycles. 
    Additionally, we show that recycled millisecond pulsars or low-mass X-ray binaries similar to those well-observed in Galactic globular clusters may also be plausible channels, but only if their duty cycle for producing bursts similar to the M81 FRB is small.
    \vspace{1cm}
\end{abstract}

\section{Introduction}

Fast radio bursts (FRBs) are bright flares of coherent radio emission with millisecond durations and large dispersion measures (DM) that imply extragalactic origin \citep[e.g.,][]{Lorimer2007,Keane2012,Thornton2013,CordesChatterjee2019}.
A fraction of FRBs have been observed to repeat \citep[e.g.,][]{Spitler2016,CHIME2019b}, indicating catalysmic models can be ruled out for at least a subset of FRBs. Perhaps the most popular model for repeating FRBs is bursts generated by young flaring magnetars \citep[e.g.,][]{PopovPostnov2013,Kulkarni2014,LuKumar2016,Metzger2017}. The recent detection of an FRB-like burst coincident with the Galactic magnetar SGR 1935+2154 \citep{CHIME2020,Bochenek2020} provides clear evidence that at least some FRBs are magnetar-powered. Magnetars are traditionally expected to form in association with massive stellar evolution, for instance in connection with standard core-collapse supernovae (SNe), superluminous SNe, and/or long gamma-ray bursts \citep[e.g.,][]{Nicholl2017}, in part due to the association of a number of Galactic magnetars with SN remnants \citep[e.g.,][]{Kaspi2017}.

The recently discovered repeating FRB 20200120E \citep{Bhardwaj2021} was localized to an old ($t\sim 10\,$Gyr) globular cluster (GC) in M81 \citep{Kirsten2021}. In a present-day GC, neutron stars (NSs)/magnetars formed in association with massive star evolution have been inactive for billions of years. Thus, standard NS progenitor models cannot explain this source.

Due to their high stellar densities, GCs are efficient factories of various high-energy sources including X-ray binaries \citep[e.g.,][]{Katz1975,Clark1975}, millisecond pulsars \citep[MSPs; e.g.,][]{Ransom2008}, cataclysmic variables \citep[e.g.,][]{Grindlay1995}, and gravitational wave (GW) sources \citep[e.g.,][]{Rodriguez2016a}. Recent work shows the central regions of core-collapsed GCs are dominated by dynamically-active massive white dwarfs (WDs) and NSs \citep[e.g.,][]{Ye2018,Kremer2020,Vitral2021,Rui2021b,Kremer2021}.
In addition to the well-observed MSP population in GCs, four young pulsars (with estimated ages $\lesssim 100\,$Myr)
are observed in Galactic GCs, suggesting that young NSs are formed at late times in GCs through some mechanism \citep[e.g.,][]{Boyles2011,Tauris2013}. Binary WD mergers \citep[e.g.,][]{King2001,Schwab2016,Kremer2021}, accretion induced collapse (AIC) of WDs in binaries \citep[e.g.,][]{Nomoto1991,Tauris2013}, binary NS mergers \citep[e.g.,][]{Rosswog2003,GiacomazzoPerna2013}, and/or NS--WD mergers \citep[e.g.,][]{Liu2018,ZhongDai2020} may all lead to young NS formation in clusters. Thus, a number of scenarios may operate in GCs that could produce a progenitor of the M81 FRB \citep{Katz2021,Lu2021}.

In this \textit{Letter}, we explore channels through which FRB emitting objects may form in GCs. In Section~\ref{sec:models}, we describe the N-body simulations used for this study. In Section~\ref{sec:progenitors}, we calculate the rate of young NS formation through various mechanisms in old GCs using a suite of N-body cluster models and compare to the inferred rate from the M81 FRB. In Section~\ref{sec:energetics}, we discuss energetics and explore whether the properties of the M81 FRB can be reasonably reproduced by these young NSs. In Section~\ref{sec:MSP}, we discuss the specific case of millisecond pulsars and X-ray binaries. We summarize our results and conclude in Section~\ref{sec:discussion}.

\begin{deluxetable}{l|c|cc}
\tabletypesize{\footnotesize}
\tablewidth{0pt}
\tablecaption{GC properties adopted in this study \label{table:GC}}
\tablehead{
	\colhead{} &
	\colhead{Total number} &
	\multicolumn{2}{c}{Number density [$\rm{pc}^{-3}$]} \\
	\colhead{} &
	\colhead{} &
	\colhead{$r<2\,$pc} &
	\colhead{$r<0.1\,$pc}
}
\startdata
Main Sequence Stars & $2.2\times10^5$ & 1000 & $10^5$ \\
\hline
Giants & 600 & 7 & $4000$\\
\hline
All White Dwarfs & $3.7\times10^4$ & 500 & $4\times10^5$\\
He White Dwarfs & 300 & 4 & $2000$\\
CO White Dwarfs & $3.6\times10^4$ & 460 & $3\times10^5$\\
ONe White Dwarfs & 1300 & 30 & $10^5$\\
\hline
Neutron Stars & 200 & 5 & $1.5\times10^4$\\
\hline
Black Holes & 0 & - & - \\
\enddata
\tablecomments{Properties for a typical core-collapsed GC with present-day total mass of $10^5\,M_{\odot}$, central velocity dispersion of $10\,\rm{km\,s}^{-1}$, and metallicity of $0.01 Z_{\odot}$. We show total number of various stellar types in the cluster and the number density within both $2\,$pc (typical half-light radius) and $0.1\,$pc (representative of the innermost region where most massive WDs/NSs reside). In core-collapsed clusters, we expect all stellar BHs have been dynamically ejected. For further details, see \citet{Kremer2021}.}
\end{deluxetable}

\begin{deluxetable*}{l|ccc|c}
\tabletypesize{\footnotesize}
\tablewidth{0pt}
\tablecaption{Formation rates for several proposed FRB progenitors \label{table:rates} in GCs}
\tablehead{
	\colhead{Event type} &
	\colhead{Total \# in models} &
	\colhead{Rate per CC GC} &
	\colhead{Volumetric rate} &
	\colhead{Active lifetime required ($\tau$)} \\
	\colhead{} &
	\colhead{} &
	\colhead{[$\rm{yr}^{-1}$]} &
	\colhead{[$\rm{Gpc}^{-3}\rm{yr}^{-1}$]} &
	\colhead{[$\times(f_v\zeta)^{-1}$]}
}
\startdata
Super-Chandrasekhar WD+WD mergers & 283 & $6\times10^{-9}$ & $4$ & $10^6\,$yr\\
(estimate including tidal capture) & - & $7\times10^{-8}$ & $45$& $10^5\,$yr\\
\hline
WD+NS mergers & 59 & $10^{-9}$ & $0.8$ & $6\times10^6\,$yr\\
(estimate including tidal capture) & - & $10^{-8}$ & $6$ & $8\times10^5\,$yr\\
\hline
NS+NS mergers & 6 & $10^{-10}$ & $0.08$ & $6\times10^7\,$yr\\
\hline
AIC from binary RLO & 21 & $5\times10^{-10}$ & $0.3$ & $2\times10^7\,$yr\\
\hline
WD+MS collisions ($M_{\rm{WD}}>1.2\,M_{\odot}$) & 1098 & $2\times10^{-8}$ & $15$ & $3\times10^5\,$yr \\
NS+MS collisions & 301 & $7\times10^{-9}$ & $4$ & $10^6\,$yr\\
\hline
\hline
Inferred rate for M81 FRB & - & - & $\approx5\times10^6/\tau$ & -
\enddata
\tablecomments{Rates for a number of events occurring in dense star clusters that may produce objects that could power FRBs. We show the rate per core-collapsed GC and the inferred volumetric rate in the local universe (assuming a GC number density of $3\,\rm{Mpc}^{-3}$ and assuming that $20\%$ of GCs have undergone core collapse, consistent with core-collapsed fraction in the Milky Way).
These rates may be viewed as upper limits, as the exact fraction of these events leading to NS formation is uncertain.
In the final column, we show the active lifetime required in order to reproduce the inferred rate of the M81 burst (scaled by the duty cycle $\zeta$ and visibility fraction $f_v$).}
\end{deluxetable*}

\section{Globular Cluster models}
\label{sec:models}

To model GCs, we use the N-body simulations from \citet{Kremer2021} computed with the cluster dynamics code \texttt{CMC} \citep{Kremer2020,Rodriguez2021}. \texttt{CMC} is a star-by-star H\'{e}non-type \citep{Henon1971a} Monte Carlo integrator that includes all relevant physics for modeling the formation/evolution of compact objects in dense stellar clusters including two-body relaxation, direct integration of 3/4-body resonant encounters, and cluster evolution within a galactic tidal field. \texttt{CMC} uses the \texttt{COSMIC} population synthesis code \citep{Breivik2020} to model stellar and binary evolution, which includes prescriptions for formation of MSPs \citep{Ye2018}.

The cluster models of \citet{Kremer2021} (18 models total) are tuned specifically to core-collapsed GCs. As shown in \citet{Ye2018,Ye2020} and \citet{Kremer2020,Kremer2021}, core-collapsed clusters are ideal environments for dynamical interactions of WDs and NSs at late times.
The most obvious reason is that core-collapsed clusters generally have the highest central densities \citep[e.g.,][]{Harris1996}. Also, because the process of cluster core collapse is connected to the dynamical depletion of stellar-mass black holes \citep[BHs; e.g.,][]{Kremer2018b}, massive WDs and NSs are expected to be the most massive objects in core-collapsed clusters, enabling WDs/NSs to efficiently mass-segregate to the cluster center \citep[e.g.,][]{Kremer2020}. Observational evidence suggests core-collapsed clusters host dense subsystems of massive WDs and NSs in their central regions \citep{Vitral2021,Rui2021b,Kremer2021}. In Table \ref{table:GC}, we list basic properties of various stellar populations in our cluster models.

To estimate the rate of a given event in old GCs, we use the following procedure. First, we count the total number of occurrences of the given event in our models at late times ($t > 9\,$Gyr) representative of the ages of the GCs in the Milky Way \citep[e.g.,][]{Harris1996} and also consistent with the age estimates for the M81 FRB's host cluster \citep{Kirsten2021}. The total rate per core-collapsed cluster can then be estimated simply as the total number of events divided by $\Delta t=5\,$Gyr (and divided by the total number of models). Since these models are roughly half the mass of typical GCs in the Milky Way, we multiply by an additional factor of two \citep{Kremer2021}. To estimate the local universe volumetric rate, we use an average volumetric number density of clusters of $3\,\rm{Mpc}^{-3}$ \citep[e.g.,][]{PortegiesZwart2000,Rodriguez2015a} and assume that roughly $20\%$ of GCs have undergone core collapse, consistent with the fraction in the Milky Way \citep[e.g.,][]{Harris1996}. Note these last two factors are uncertain (although not likely by more than a small factor). Thus, our model-inferred rates can be viewed with a factor of a few uncertainty.

M81 is expected to host roughly 200-500 GCs \citep{PerelmuterRacine1995,Chandar2004} (a factor of a few more than the Milky Way) with mean metallicity $[\rm{Fe/H}]=-1.48\pm0.19$ \citep{Perelmuter1995} (comparable to the Galactic GCs). Thus the \citet{Kremer2021} models reasonably capture the properties of the M81 GCs.

\section{Young neutron stars formation scenarios}
\label{sec:progenitors}

The single repeating FRB in M81 detected at a distance of $3.6\,$Mpc implies a volumetric density of at least $n_{\rm{FRB}}\approx5\times10^6\,\rm{Gpc}^{-3}$.
The formation rate for the source of the M81 FRB can then be inferred as
\begin{multline}
    \label{eq:inferred_rate}
    \mathcal{R}_{\rm{src}} \approx 50 \Bigg(\frac{n_{\rm{FRB}}}{5\times10^6\,\rm{Gpc}^{-3}} \Bigg) \\ \times
    \Bigg(\frac{\tau}{10^5\,\rm{yr}} \Bigg)^{-1} \Bigg(\frac{1}{f_v \zeta} \Bigg) \rm{Gpc}^{-3}\,\rm{yr}^{-1},
\end{multline}
where $\tau$ is the lifetime of the FRB source, $f_v$ is the visibility fraction (the chance of seeing a burst from the source, which is related to the beaming factor, luminosity function, etc.), and $\zeta$ is the duty cycle (the fraction of time a source is actively producing bursts similar to the M81 FRB during its lifetime). Any viable formation channel for producing the M81 FRB must create FRB sources at a rate comparable to $\mathcal{R}_{\rm{src}}$.

In the following subsections, we summarize each of the FRB progenitor channels that we consider in this work. As a reference, their rates that we calculate using N-body simulations are summarized in Table \ref{table:rates}. Dividing the volumetric rate by $n_{\rm{FRB}}$, we derive the active FRB lifetime required for each channel to reproduce the M81 FRB (summarized in the last column of Table~\ref{table:rates}). This timescale is an important constraint on the properties of the NSs produced from any of these channels.

\subsection{White Dwarf Mergers}
\label{sec:WDWD}

We first discuss the case where a young NS is formed through a double WD merger. Our theoretical understanding of the outcomes of WD mergers has developed considerably in recent years, but there are still many uncertainties. For pairs of CO WDs, the merger may promptly trigger detonation of the more massive WD, producing a Type Ia SN \citep[e.g.,][]{Shen2018,Perets2019}. Alternatively, if the merger is non-destructive, a CO-dominated remnant results \citep{Schwab2021}. This further evolves over a timescale of $\sim10\,{\rm kyr}$, first burning to produce a remnant with an ONe composition, and then, if this exceeds the Chandrasekhar mass \citep[depending on the highly uncertain mass-loss rate during the puffed-up CO giant phase; e.g.,][]{Yoon2007}, eventually succumbing to electron capture to collapse to a NS \citep{Nomoto1985,Saio1985,Schwab2016}. This is sometimes referred to as merger induced collapse (MIC). If one or more of the merging WDs is ONe composition to begin with (and if the total merger mass exceeds the Chandrasekhar mass), then a detonation will be prevented and MIC will almost certainly occur \citep[e.g.,][]{Nomoto1991}.

The properties of NSs formed via MIC can be estimated from simple arguments. Immediately post merger, the remnant has significant angular momentum taken from the compact orbit of $\sim10^{50}\,{\rm erg\,s}$. During the subsequent evolution, there is significant angular momentum loss from a combination of viscous evolution, inflation during the burning processes, and mass shedding, resulting in a remnant angular momentum of $\sim10^{48}\,{\rm erg\, s}$ \citep{Shen2012,Schwab2012,Schwab2021}. With the eventual MIC, the resulting NS would have a spin period of $\sim10\,{\rm ms}$. The magnetic field strength is more uncertain. The hot, differentially-rotating merger remnant may generate strong fields, and \citet{GarciaBerro2012} show that a dynamo can easily produce fields of $\sim10^7\,{\rm G}$. Flux conservation during collapse to a NS would amplify this value by $\sim10^4$, leading to a field of $\sim10^{11}-10^{12}\,{\rm G}$. We note that this more recent picture for MIC differs from previous discussions where it is assumed that MIC leads to a millisecond magnetar \citep{Usov1992,King2001,Levan2006}. Nevertheless, as we show in this work, the longer lifetime of a NS with a longer spin period and weaker magnetic field may in fact be more consistent with the active lifetime needed to explain the rate of FRBs like the one in M81.

In total, we identify 305 binary WD mergers in our cluster models at late times. Of these, 283 (roughly $93\%$) have a total mass in excess of the Chandrasekhar limit and 177 (roughly $58\%$) have at least one ONe component. Assuming all of these super-Chandrasekhar mergers lead to MIC and NS formation, we estimate young NSs are formed through WD mergers at a rate of roughly $6\times10^{-9}\,\rm{yr}^{-1}$ per typical core-collapsed cluster and estimate a volumetric rate of $4\,\rm{Gpc}^{-3}\,\rm{yr}^{-1}$ in the local universe. We also record in \texttt{CMC} the number of direct physical collisions of WD pairs. As pointed out in \citet{Kremer2021}, the pericenter distance necessary for a pair of WDs to be \textit{tidally captured} (either during a single--single encounter or during a binary resonant encounter) may be a few$-10$ times larger than the distance required for physical collision \citep[the WD radius; e.g.,][]{Samsing2017}. Although direct collisions themselves may be more likely to lead to an explosive transient outcome \citep[e.g.,][]{Katz2012}, knowing the total number of direct collisions enables us to estimate in post-processing the rate of binary WD mergers that may occur through tidal capture \citep[as discussed in][GW inspiral and merger is the most likely ultimate outcome of tidal capture]{Samsing2017}. We identify 403 WD--WD collisions in total (roughly $87\%$ of which are super-Chandrasekhar), implying a WD merger rate through tidal capture of up to roughly $7\times10^{-8}\,\rm{yr}^{-1}$ per typical core-collapsed GC and a volumetric rate of roughly $45\,\rm{Gpc}^{-3}\,\rm{yr}^{-1}$ in the local universe.

\subsection{White Dwarf--Neutron Star Mergers}

A number of studies argue WD--NS mergers could lead to spun-up NS remnants, possibly with ultra-strong magnetic fields \citep[e.g.,][]{Paschalidis2011a,MargalitMetzger2016,Liu2018,KhokhriakovaPopov2019}.
In this case, \citet{ZhongDai2020} pointed out that flaring magnetized NSs formed from NS--WD mergers may have burst energetics and host galaxy properties consistent with FRBs similar to FRB 180924 \citep{Bannister2019}.

In total, we find 59 NS--WD mergers in our cluster models.
We estimate a WD--NS merger rate of $10^{-9}\,\rm{yr}^{-1}$ per typical GC and a volumetric rate of roughly $0.8\,\rm{Gpc}^{-3}\,\rm{yr}^{-1}$.
As in the WD--WD merger case, the rate of NS--WD mergers may also increase significantly if tidal capture were incorporated \citep{Samsing2017}. Assuming again the cross section for tidal capture is up to roughly a factor of ten larger than the cross section for physical collision (we identify 37 WD--NS collisions in our models), the total rate of WD--NS mergers may increase to roughly $6\,\rm{Gpc}^{-3}\,\rm{yr}^{-1}$ in the local universe.

\subsection{Neutron Star Mergers}

A massive magnetized NS remnant is the expected outcome of a binary NS merger driven by GW inspiral \citep[e.g.,][]{Rosswog2003, GiacomazzoPerna2013}. Generally, this remnant will be well above the Tolman-Oppenheimer-Volkoff (TOV) mass and will be stable only temporarily before collapse into a BH. However, depending on various uncertain features,
a subset of NS mergers may produce long-lived NS remnants \citep[e.g,][]{Piro2017,MargalitMetzger2019,BeniaminiLu2021}. \citet{Margalit2019} demonstrated that magnetars born from binary NS mergers may account for a subset of observed FRBs, notably FRB 180924 whose host galaxy and offset match well the observed distributions for short gamma-ray bursts \citep[e.g.,][]{Berger2014}. Magnetic interactions may also produce sufficient energy to power FRBs in the seconds \citep{Piro2012} or even centuries \citep{Zhang2020} prior to merger.

In our models, we identify only six binary NS mergers at late times.
This translates to a BNS merger rate of $10^{-10}\,\rm{yr}^{-1}$ per typical core-collapsed GC and a volumetric rate of roughly $0.08\,\rm{Gpc}^{-3}\,\rm{yr}^{-1}$ in the local universe, consistent with the rates predicted in \citet{Ye2020}. Thus, even in the extremely optimistic case that \textit{all} of these NS mergers lead to long-lived NS remnants, binary NS mergers contribute negligibly to the production of young NSs at late times relative to other channels.

\subsection{Accretion Induced Collapse from Binary Mass Transfer}

Another commonly discussed scenario for NS formation is AIC of a massive WD initiated by mass transfer from a binary companion such as a main sequence (MS) star, red giant, or helium star/WD \citep[e.g.,][]{Nomoto1991,Tauris2013,Schwab2015}. Similar to the WD merger case, when an ONe WD accretes to the Chandrasekar limit, rapid electron-captures onto heavy elements produced by the oxygen burning is expected to prevent a thermonuclear explosion \citep[e.g.,][]{Miyaji1980}. Thus, NS formation is generally considered the most likely outcome of AIC of a WD, though see \citet{Jones2016,Jones2019} for discussion of alternative scenarios where a thermonuclear explosion results.

In GCs, binaries with ONe WDs can come from the primordial binary population, dynamical exchange encounters, and/or tidal capture. These binaries may be hardened to Roche lobe overflow through subsequent dynamical encounters \citep[e.g.,][]{Ivanova2008,Belloni2016,Ye2018,Kremer2021}.
Here, we follow the methods outlined in \citet{Ye2018} \citep[which in turn follow the presciptions of][]{Hurley2002} to treat AIC from binary mass transfer. In total, we identify 21 NSs formed at late times through this process.
In total, we estimate that young NSs are formed at a rate of roughly $5\times10^{-10}\,\rm{yr}^{-1}$ per typical core-collapsed GC and a volumetric rate of roughly $0.3\,\rm{Gpc}^{-3}\,\rm{yr}^{-1}$.

\subsection{Alternative formation channels}

Here, we discuss two additional scenarios involving the collisions of WDs/NSs with MS stars.
These collisions occur when the pericenter distance of a pair of stars is less than the sum of the two stellar radii, $r_p < R_1+R_2$. This criterion may be met during both single--single encounters and binary-mediated resonant encounters. In GCs, the colliding MS star is most commonly an M-dwarf ($\sim0.5\,M_{\odot}$), due simply to the expected mass function in old clusters \citep{Kremer2019c}. 
Following disruption of the star, a thick super-Eddington disk likely forms \citep[e.g.,][]{Kremer2019c}. In the adiabatic inflow-outflow model of \citet{BlandfordBegelman1999}, only a fraction, $10(r_{\rm{a}}/r_{\rm{d}})$ (where $r_{\rm{d}}\approx R_{\star}$ is the disk radius and $r_{\rm{a}}$ is the radius of the accretor), of the mass supplied at large radii is likely accreted onto the compact object.

For NS--MS collisions, we estimate a factor of $10^{-4}$ of the available material ($\lesssim0.5\,M_{\odot}$ for an M-dwarf) may be accreted. Although unlikely to grow the NS significantly, this accreted material may be sufficient to spin up the NS, potentially to millisecond periods.
In this case, the formation of a rapidly-spinning NS is plausible. In total, we identify 301 NS--MS collisions in our cluster models, translating to a rate of roughly $7\times10^{-9}\,\rm{yr}^{-1}$ per typical core-collapsed GC and a volumetric rate of roughly $4\,\rm{Gpc}^{-3}\,\rm{yr}^{-1}$.

In the WD--MS collision case,
a larger fraction of available mass may be accreted. Thus it is plausible \textit{massive} WDs may accrete to the Chandrasekhar limit and collapse. In total, we identify 4750 WD--MS collisions in our models. Of these 1098 (475) involve a WD of mass in excess of $1.2\,\,(1.3)\,M_{\odot}$. Assuming as an upper limit these all lead to AIC, this yields a NS formation rate of roughly $2\times10^{-8}\,\, (10^{-8})\,\rm{yr}^{-1}$ per typical core-collapsed cluster and a volumetric rate of roughly $15\,\,(6)\,\rm{Gpc}^{-3}\,\rm{yr}^{-1}$ in the local universe. Many aspects of this process are of course highly uncertain. For example, we have not considered the possibility that nuclear reactions ignited on the WD surface may supply sufficient energy to unbind the disrupted stellar material completely \citep[e.g.,][]{SharaShaviv1977}. More detailed simulations are ultimately necessary to predict more precisely the outcome of WD--MS collisions and determine whether AIC is a plausible outcome, and if so the properties of the NS that may form.

\section{Discussion}
\label{sec:energetics}

\begin{figure*}
    \begin{center}
    \includegraphics[width=0.9\linewidth]{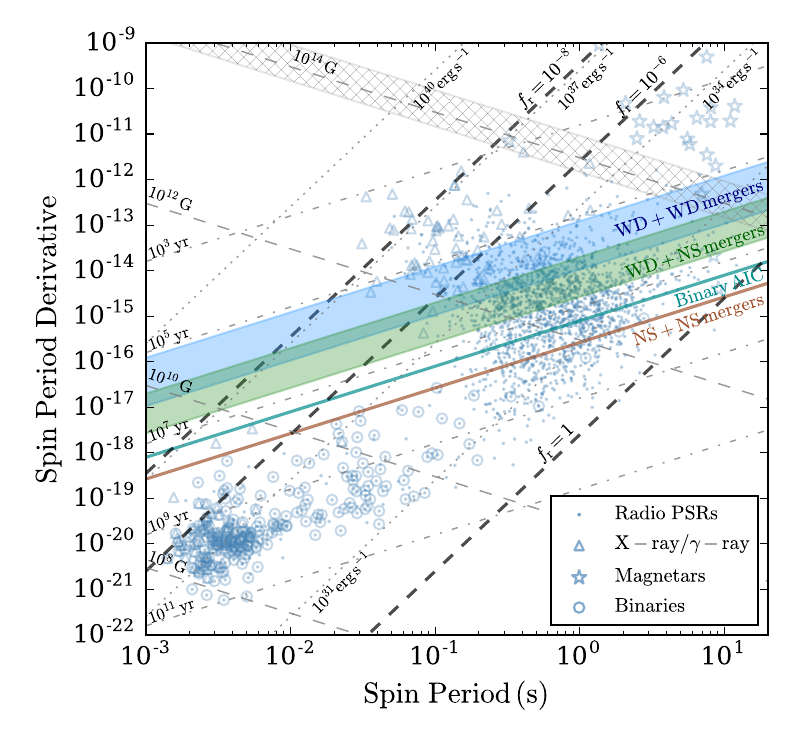}
    \caption{\footnotesize  $P \dot{P}$ diagram. Overlaid colored curves denote the active FRB lifetime $\tau$ required to reproduce the inferred volumetric event rate inferred for the M81 FRB ($5\times10^6/\tau\,\rm{Gpc}^{-3}\,\rm{yr}^{-1}$). Different colors denote the various FRB progenitor formation channels in GCs discussed in Section~\ref{sec:progenitors}.
    The dashed black curves show the isotropic equivalent time-averaged luminosity for the M81 FRB assuming efficiency factors of $f_{\rm{r}}=10^{-8}$, $f_{\rm{r}}=10^{-6}$, and $f_{\rm{r}}=1$ for creation of coherent radio emission (see text for details). The hatched gray band denotes the allowable parameter space for a magnetically-powered source with $f_{\rm{r}}\gtrsim 10^{-4}$. For reference, we include in blue all radio sources in the ATNF Pulsar Catalog \citep{Manchester2005} and magnetars from the McGill Magnetar Catalog \citep{OlausenKaspi2014}.}
    \label{fig:PPdot}
    \end{center}
\end{figure*}

By adding the total burst fluence of $6.6\,\rm{Jy\,ms}$ for the M81 FRB \citep[Table 1 of][]{Bhardwaj2021} over CHIME's total on-source time of $\approx100\,$hr, we estimate a time-averaged isotropic equivalent luminosity of $\langle \dot{E} \rangle \approx10^{29}f_{\rm{r}}^{-1}\,\rm{erg\,s}^{-1}$.\footnote{
If the spectral index of the luminosity function is greater than 2, the radio flux would be dominated by faint undetectable bursts. Therefore, without knowing the luminosity function, this estimate of the average luminosity should be considered a lower limit. Hence, all estimates of radio efficiency are effectively lower limits as well. 
} Here $f_{\rm{r}}$ is the highly-uncertain efficiency factor for creating coherent radio emission. 
The total energy required to supply the bursting activity for time $\tau \approx10^5\,$yr is then $E_{\rm{FRB,tot}} \approx 5\times10^{41}f_{\rm{r}}^{-1}\,\rm{erg}$. Energy emission from NSs can be broadly divided into two categories: magnetically powered and rotation powered. We consider each in turn next.

In the magnetically-powered scenario, the intrinsic energy budget is estimated as $E_{\rm{mag}} \approx (B_{\rm{NS}}^2R_{\rm{NS}}^3$)/6.
As in \citet{BeloborodovLi2016,Margalit2019}, we estimate the magnetic activity timescale in the modified Urca case as
\begin{equation}
    \label{eq:t_mag}
    \tau_{\rm{mag}} \approx 2\times 10^5 \Bigg( \frac{B_{\rm{NS}}}{10^{14}\,\rm{G}} \Bigg)^{-1.2}
    \Bigg(\frac{L}{10^{5}\,\rm{cm}} \Bigg)^{1.6} \rm{yr},
\end{equation}
assuming $\delta B \sim B/2$ as the amplitude of magnetic field fluctuations over an (uncertain) length-scale $L$ within the NS core. In this case, the characteristic luminosity from magnetic activity is
\begin{equation}
    \label{eq:Edot_mag}
    \langle \dot{E}_{\rm{mag}} \rangle \approx E_{\rm{mag}}/\tau_{\rm{mag}} \approx 3\times10^{32} \Bigg( \frac{B_{\rm{NS}}}{10^{14}\,\rm{G}} \Bigg)^{3.2} \rm{erg\,s}^{-1}.
\end{equation}
By combining Equations (\ref{eq:t_mag}) and (\ref{eq:Edot_mag}) with the inferred time-averaged luminosity, we find radio emission efficiencies of $f_{\rm{r}} \gtrsim10^{-4}$ (and field strengths of roughly $2\times10^{14}\,$G) are required to produce active lifetimes, $\tau_{\rm{mag}}\approx10^5\,$yr, necessary to explain the FRB event rate for the young NS formation scenarios discussed in Section~\ref{sec:progenitors}. Although highly uncertain, radio emission efficiencies this high 
have been demonstrated by some FRB models \citep[e.g.,][]{Plotnikov2019,LuKumarZhang2020}.

In the rotation-power scenario, the intrinsic energy budget is simply the rotational energy of the NS: $E_{\rm{rot}} \approx 2\pi^2I P_{\rm{NS}}^{-2}$
where $I\approx (2/5)MR^2$ is the moment of inertia of the NS and $P_{\rm{NS}}$ is the NS spin period.
The characteristic spin-down timescale of a magnetic dipole is

\begin{equation}
    \tau_{\rm{sd}} \approx \frac{P}{2\dot{P}} \approx 5\times 10^5\,\rm{yr} \Bigg( \frac{\it{P}_{\rm{NS}}}{10\,\rm{ms}}\Bigg)^2 \Bigg( \frac{\it{B}_{\rm{NS}}}{10^{11}\,\rm{G}}\Bigg)^{-2}, 
\end{equation}
which yields a characteristic spin-down luminosity of
\begin{equation}
    \label{eq:lum_sd}
    \dot{E}_{\rm{rot}} \approx 10^{37}  \Bigg( \frac{\it{B}_{\rm{NS}}}{10^{11}\,\rm{G}}\Bigg)^{2} \Bigg( \frac{\it{P}_{\rm{NS}}}{10\,\rm{ms}}\Bigg)^{-4} \rm{erg\,s}^{-1}.
\end{equation}
For $f_{\rm{r}}\gtrsim10^{-8}$, the spin-down luminosity is sufficient to supply the energy required to power the FRB for a $\tau_{\rm{sd}}\gtrsim10^5\,$yr lifetime. 

In Figure \ref{fig:PPdot}, we use a $P\dot{P}$ diagram to explore the range of NS properties required to produce the rates and energetics inferred for the M81 FRB. As colored lines, we show the minimum active lifetime $\tau$ required for a young NS formed through the various channels discussed in Section~\ref{sec:progenitors} to produce the volumetric density of $n_{\rm{FRB}}=5\times10^6\,\rm{Gpc}^{-3}$ inferred from the M81 FRB (assuming a duty cycle and $f_v$ of order unity). The thick blue and green bands denote the range in timescales inferred for WD--WD and WD--NS mergers, respectively, reflecting the uncertainty in the contribution of tidal capture to the merger rate (as discussed in Section~\ref{sec:progenitors}). Dark cyan and red curves show the inferred rates for AIC in binary systems and NS--NS mergers, respectively.
As dashed black curves, we show the time-averaged isotropic equivalent luminosity for three different radio emission efficiency factors that bracket the large uncertainty in this parameter: $f_{\rm{r}}=1$ (as an unrealistic upper limit), $f_{\rm{r}}=10^{-6}$ \citep[representative of the efficiency inferred for the Galactic magnetar;][]{CHIME2020}, and $f_{\rm{r}}=10^{-8}$ \citep[representative of highly inefficient radio emission as adopted in some previous studies; e.g.,][]{Lyubarsky2014,Metzger2017}. For reference, we also show as blue markers Galactic magnetars (blue stars), X-ray/$\gamma$-ray emitters (triangles), and standard radio pulsars (blue points; open circles indicate pulsars found in binary systems). See \citet{Manchester2005} and \citet{OlausenKaspi2014} for further detail.

As a hatched gray band, we show the range in magnetic field values required for a magnetically-powered energy source for radio efficiencies ranging from $f_{\rm{r}}\approx10^{-4}$ (minimum required value corresponding to $B\approx2\times10^{14}\,$G and $\tau_{\rm{mag}}\approx10^5\,$yr) to $f_{\rm{r}}\approx10^{-3}$ (corresponding to $B\approx7\times10^{13}\,$G and $\tau_{\rm{mag}}\approx3\times10^5\,$yr). As shown, a magnetically-powered scenario would require a magnetic activity timescale in excess of the spin-down timescale (with the exception of objects with $P_{\rm{NS}}\gtrsim10\,$s, which are disfavored by theoretical models; see Section~\ref{sec:progenitors}). Furthermore, such magnetically-powered sources must be physically distinct from the Galactic magnetars \citep{DuncanThompson1992}
which have empirically constrained lifetimes $\lesssim10^4$yr \citep{Kaspi2017}, too small to explain the inferred rate of the M81 FRB for any of the young NS formation channels described here (see Table \ref{table:rates}).

\textit{The intersection between the bold dashed lines and colored lines indicate regions of parameter space that reproduce the rate and energetics of the M81 FRB for a spin-down powered source.} These regions also therefore imply allowable magnetic field ($B\propto \sqrt{P\dot{P}}$) for the M81 progenitor.
This shows that highly-magnetized NSs ($B\gtrsim5\times10^{14}\,$G) with spin-down timescales $\lesssim10^4$yr, can likely be ruled out based on rate arguments alone, as the active lifetimes of these objects are too small.
Additionally, unless NS--NS mergers produce NS remnants with lifetimes $\gtrsim10^8$yr \citep[high unlikely; e.g.,][]{Piro2017}, the binary NS merger scenario can likely also be ruled out by the rate alone.

For radio emission efficiencies $f_{\rm{r}}\lesssim10^{-6}$, WD--WD mergers, WD--NS mergers and AIC in a binary system all appear viable for the rotation-powered scenario, provided these channels can produce NSs with magnetic fields strengths in the range $\approx10^{11}-10^{12}\,$G and spin periods in the range $\approx10-100\,$ms. Intriguingly, these properties are comparable to those predicted for NS remnants formed through WD mergers (see Section~\ref{sec:WDWD}). In principle, the WD--MS and NS--MS collision scenarios are plausible based on the rates and energetics, however more detailed simulations are required to test whether indeed these events may lead to magnetized/spinning NS remnants.

The region of the $P\dot{P}$ diagram occupied by young NSs with required properties for the M81 FRB is near the region occupied by Galactic pulsars (particularly for $f_{\rm{r}}\gtrsim10^{-6}$), most of which are associated with SN remnants and therefore likely formed through core collapse.
The core-collapse SN rate \citep[roughly $10^5\,\rm{Gpc}^{-3}\,\rm{yr}^{-1}$; e.g.,][]{Taylor2014} is substantially higher than the NS formation rate in old GCs.
In this case, a factor of $\gtrsim1000$ more FRBs similar to the M81 burst would be produced by the spin-down luminosity powered pulsar population in the galactic field.
Unless there is a yet-understood mechanism that may prevent NSs in the field from emitting FRBs but that does not operate in GCs, this implies a spin-down-powered model with radio efficiency factor $\lesssim10^{-7}$ may be necessary. In this case, the NS properties required to explain the rate and energetics occupy a region of parameter space comfortably unique from any of the known Galactic radio sources.

We have assumed in Figure \ref{fig:PPdot} that the duty cycle $\zeta$ to flare as a repeating FRB similar to the M81 source is of order unity. Although the exact duty cycle is highly uncertain, this is likely an optimistic assumption. As shown in Equation (\ref{eq:inferred_rate}), lower duty cycles would require larger active lifetimes.
We have also assumed $f_v$ of order unity, also an optimistic choice. For instance some studies argue a beaming factor $f_v\approx0.1$ \citep[consistent with the value expected for pulsars;][]{TaurisManchester1998} may be more appropriate. If for instance we adopt $f_v\approx0.1$ and $\zeta \approx 0.1$ \citep[as in, e.g.,][]{Nicholl2017}, the binary AIC channel and WD--NS merger scenarios may be ruled out, as NSs with lifetimes in excess of roughly $10^8\,$yr would be required to reproduce the inferred rate, in tension with values predicted from simulations. In this case, WD--WD mergers (especially invoking the tidal capture scenario discussed in Section~\ref{sec:WDWD}) may be the only viable young NS scenario, assuming NSs with active lifetimes $\gtrsim10^6\,$yr can be produced.

We use time-average luminosity in this paper, since it gives a lower limit on the energy reservoir required to supply the bursting activity over the source lifetime.
Note that the peak luminosity of some of the bright bursts from this source \citep[$\dot{E}_{\rm{peak}}\approx 10^{39}f_{\rm{r}}^{-1}\,\rm{erg\,s}^{-1}$; ][]{21Majid} may be larger than the expected spin-down luminosities invoked to power the source ($\sim10^{37}\,\rm{erg\,s}^{-1}$). However, this can be reconciled with a Doppler beaming, where the luminosity per solid angle is boosted by the Lorentz factor $\gamma^4$ in the observed frame relative to the inertial frame. This process requires that the emission region be smaller than the relativistic beaming angle, similar to what has been invoked to explain observations of the Crab pulsar \citep{Bij2021}. With $\gamma\gtrsim10$ \citep[minimum requirement for FRBs from curvature radiation models; e.g.,][]{Kumar2017, KatzFRBSGR}, the rest frame peak luminosity will be below the spin-down luminosity. 
The time-average luminosity would average over instances when the relativistic beaming is pointed away from us.

\section{A millisecond pulsar or X-ray binary progenitor?}
\label{sec:MSP}
GCs are well-known to efficiently create MSPs \citep[e.g.,][]{Ransom2008}. MSPs
are expected to form in GCs when NS binaries (either primordial binaries or binaries assembled dynamically through exchange encounters) are driven to mass transfer through a combination of stellar evolution of the companion and hardening by repeated fly-by encounters in the cluster \citep[e.g.,][]{Pooley2003,Ivanova2008,Ye2018}. To date, roughly $200$ MSPs have been observed in 36 Galactic GCs\footnote{http://www.naic.edu/~pfreire/GCpsr.html}. In our cluster models, we identify 105 total MSPs \citep[defined as spin periods less than $30\,$ms and formed through binary mass transfer channels similar to those described in][]{Ye2018}, or roughly 6 MSPs per cluster, consistent with the $200/36\approx5$ MSPs per cluster one can infer from the observed population.

Given the well-observed abundance of MSPs in GCs, one may ask if these objects may plausibly explain the M81 FRB.
Unlike the scenarios discussed in Section~\ref{sec:progenitors}, the formation rates and spin-down timescales of MSPs in clusters are relatively well understood from the actual observed populations. Thus for the MSP scenario, we can constrain empirically the radio efficiency and duty cycle required to explain the observed FRB.

The relatively low magnetic fields and high spin periods imply low spin-down rates for MSPs and long lifetimes $\tau_{\rm{sd}}\approx10^{10}\,$yr. The spin-down luminosity of a typical MSP (assuming $P=3\,$ms and $B=5\times 10^8\,$G) is roughly $4\times10^{34}\,\rm{erg\,s}^{-1}$ (Equation \ref{eq:lum_sd}). Thus, the roughly $10^{29}f_{\rm{r}}^{-1}\,\rm{erg\,s}^{-1}$ time-averaged luminosity inferred for the M81 FRB can be reproduced for radio emission efficiencies $f_{\rm{r}} \gtrsim10^{-6}$. From a rates perspective, assuming roughly 5-10 MSPs per cluster, we can estimate an MSP volumetric density of roughly $10^{10}\,\rm{Gpc}^{-3}$ in GCs in the local universe. Thus, MSPs could explain the density of repeating FRBs inferred from the M81 burst ($\approx5\times10^6\,\rm{Gpc}^{-3}$) if the duty cycle for MSPs to produce repeating FRBs similar to the observed burst is roughly $10^{-4}$. Thus (pending the uncertain details of radio efficiency and duty cycle which future studies may elucidate), we conclude a MSP could reasonably explain the M81 FRB. 

Recent studies have suggested that accretion-powered compact objects could also be viable (repeating) FRB sources, for instance as generated by plasmoids ejected from the accretion funnel \citep{Katz2017,Sridhar2021,Deng2021}. In particular, NS X-ray binaries \citep[the expected progenitors of MSPs; e.g.,][]{Alpar1982} may be relevant in the GC context, as these systems are observed in abundance in GCs \citep[e.g.,][]{Clark1975,Heinke2010}.\footnote{As discussed in \citet{Sridhar2021}, BH X-ray binaries are also viable sources. Unlike their NS counterparts, only a handful of accreting BH binary candidates have been observed in GCs \citep[e.g.,][]{Strader2012}. Since NS X-ray binaries are expected to be more numerous, we focus only on these sources here. Future work may consider the possibility of FRBs generated by accreting stellar-mass BHs in GCs.} Theory and observations \citep[e.g.,][]{Heinke2003,Ivanova2008} suggest an average of $\sim1-10$ NS X-ray binaries per typical GC (in our models, we find roughly 1 accreting NS binary per model at late times), implying a volumetric density $\gtrsim10^9\,\rm{Gpc}^{-3}$ in clusters.
Thus, similar to the MSP scenario, X-ray binaries are likely only viable progenitors if their duty cycles for bursts similar to the M81 FRB are small.

\section{Summary and Conclusions}
\label{sec:discussion}

In old GCs, various dynamical scenarios create NSs that may plausibly power FRBs similar to the repeating FRB observed in M81. Using a suite of N-body cluster models, we have shown that WD--WD mergers, WD--NS mergers, and AIC of WDs in binary systems are all plausible candidates, assuming visibility fraction and duty cycles for repeating FRB emission of order unity. \textit{For less optimistic choices of visibility fraction and duty cycle, WD--WD mergers are likely the only scenario that may yield a sufficiently high rate.}

We consider two energy emission mechanisms. We show a magnetically-powered source (e.g., a magnetar) may be viable for $f_{\rm{r}}\gtrsim10^{-4}$. These objects would have to be distinct from the magnetars observed in the Milky Way, which have empirically constrained lifetimes $\lesssim10^4\,$yr, too short to explain the inferred event rate of the M81 FRB. Additionally, the magnetic activity lifetimes of these magnetically-powered objects would need to exceed their spin-down lifetimes. Alternatively, if magnetic fields strengths of $\approx10^{11}\,$G and spin periods of $\approx10\,$ms can be produced \citep[consistent with those predicted from WD merger models; e.g.,][]{Schwab2021}, \textit{rotation-powered} NS remnants formed through these scenarios can plausibly explain both the rate and burst energetics inferred from the M81 FRB. WD--MS collisions and NS--MS collisions occur at a high enough rates to also be viable progenitors, although whether or not a rapidly-spinning and/or highly-magnetized NS may be produced in these scenarios is less certain. The relatively low event rate of NS--NS mergers implies this channel is a less likely scenario.
In addition to the young NS scenario, we also showed that recycled MSPs with spin-down times of $\gtrsim10^{10}\,$yr as well as X-ray binaries may be viable channels from a rates and energetics perspective.

If indeed MSPs and/or young NSs formed through the AIC/MIC of WDs provide a channel for FRBs in GCs, this FRB channel should operate similarly for analogous sources in the galactic field.
The rate of AIC in the galactic field (through either stable accretion from a companion star or through the merger of a super-Chandrasekhar WD binary) is highly uncertain, with previous studies predicting rates in the range $0.1-10^2\,\rm{Gpc}^{-3}\,\rm{yr}^{-1}$ \citep[e.g.,][]{YungelsonLivio1998,Fryer1999, Tauris2013,Kwiatkowski2015}.

Additionally, the number of X-ray binaries per unit stellar mass is roughly a factor of 100-1000 times higher in GCs compared to the galactic field \citep[e.g.,][]{Clark1975}. In the Milky Way, GCs constitute roughly $0.1\%$ of the total Galactic stellar mass. This implies roughly a factor of 10 more MSPs/X-ray binaries (and therefore potential FRBs) in the field compared to clusters. 
The M81 FRB has the lowest extragalactic dispersion measure (DM) in the CHIME/FRB Collaboration catalog \citep{Bhardwaj2021,CHIME2021}. The low DM is critical to constraining the source distance and host association. Being offset from their host galaxies, GCs contribute relatively little host DM compared to galactic field environments. Therefore, roughly 10 more FRBs from galactic field environments at similar distances to M81 may simply have not yet been localized because of the relatively large host DM contribution. As more FRBs are localized to the fields and clusters of galaxies in the future, it will provide a better test of the progenitor scenario.

We have focused specifically on the possibility that the M81 FRB occurred in a core-collapsed GC. In non-core-collapsed clusters, the rates of the various formation scenarios summarized in Table \ref{table:rates}, are roughly 10-100 times smaller \citep[see, e.g.,][]{Kremer2020}, due to the relatively low central densities facilitated by the presence of large numbers of stellar-mass BHs. Current observations place rough constraints on the total mass and metallicity of the host cluster for the M81 FRB (both consistent with the models adopted in this study), however detailed features of the host (such as the density profile, core radius, etc) are not constrained. Future observations may further constrain the FRB's host cluster, including whether or not the cluster is core collapsed.

\acknowledgements{
We thank Sterl Phinney for discussions during the early stages of this work, Josiah Schwab for helpful feedback on the fate of white dwarf mergers and comments on the manuscript, Wenbin Lu, Ben Margalit, and Bing Zhang for useful discussions, and the anonymous referee for constructive comments and suggestions.
KK is supported by an NSF Astronomy and Astrophysics Postdoctoral Fellowship under award AST-2001751.
}

\bibliographystyle{aasjournal}
\bibliography{mybib}

\end{document}